\documentclass[preprint]{aastex}
\usepackage{colordvi}
\usepackage{color}
\usepackage[hyphens]{url}

\begin{document}

\title{Strong Transverse Photosphere Magnetic Fields and Twist in Light Bridge Dividing $\delta$ Sunspot of Active Region 12673 }

\author{Haimin Wang\altaffilmark{1},
        Vasyl Yurchyshyn\altaffilmark{1},
        Chang Liu\altaffilmark{1},
        Kwangsu Ahn\altaffilmark{1},
        Shin Toriumi \altaffilmark{2}
        and Wenda Cao\altaffilmark{1}}

\affil{1.\ Big Bear Solar Observatory, \\ New Jersey Institute of Technology \\
40386 North Shore Lane, Big Bear City, CA 92314}

\affil{2. \ NAOJ Fellow, National Astronomical Observatory of Japan, \\2-21-1 Osawa, Mitaka, Tokyo 181-8588, Japan}


Solar Active Region (AR) 12673 is the most flare productive AR in the solar cycle 24. It produced four X-class flares including the X9.3 flare on 06 September 2017 and the X8.2 limb event on 10 September. Sun \& Norton (2017) reported that this region had an unusual high rate of flux emergence,  while Huang et al. (2018) reported that the X9.3 flare had extremely strong white-light flare emissions.  Yang at al. (2017) described the detailed morphological evolution of this AR. In this study,  we focus on usual behaviors of the light bridge (LB) dividing the $\delta$ configuration of this AR, namely the strong magnetic fields in the LB and  apparent photospheric twist as shown in observations with a 0.1\arcsec\ spatial resolution obtained by the 1.6~m GST (Cao et al. 2010) at BBSO.

Both vector magnetograms from the spectropolarimeter (SP) of Hinode's Solar Optical Telescope (SOT) and the Helioseismic and Magnetic Imager (HMI) on board the Solar Dynamics Observatory use Stokes inversion with the maximum magnetic field strength limited to 5000~G. This limitation is valid for the vast majority of solar ARs.  In previous studies,  the highest magnetic field was reported to be 4300~G at the magnetic polarity inversion line (PIL) dividing $\delta$ sunspots (Zirin \& Wang 1993), where the strong field is parallel to the solar surface.  Very recently,  Okamoto \& Sakurai (2018) reported strong fields as high as 6250~G, also at the PIL region.  Viewing Hinode/SP and SDO/HMI magnetograms of the AR 12673,  we noted many pixels along the flaring PIL are saturated at the maximum value of 5000~G.  This motivated us to examine Stokes profiles recorded by the Near Infra-Red Imaging Spectropolarimeter (NIRIS) of GST -- a Fabry-P{\'e}rot based imaging Stokes polarimeter that provides scanning of 60 line positions of the Fe~{\sc i} 1.56$\mu m$ line  (Wang et al., 2017). With the large Lande factor ($g$=3.0) at long wavelength, the Zeeman splitting is large.

In Figures~1(a) and (b),  we show the line-of-sight (LOS) and transverse magnetic field strength using the Hinode/SP level 2 data. They are obtained a few hours after the X9.3 flare on 06 September 2017.  While the LOS fields are under 5000~G, the transverse fields have several patches at the saturation level of 5000~G. These patches are concentrated at the two PIL segments of the $\delta$ configuration, as outlined by the two white boxes.  To investigate the properties of these two areas,  we study the corresponding  photospheric structures with BBSO/GST observations in the TiO band, a good proxy for the photospheric continuum. It is obvious from Figure~1(c) that LBs in these two sections of strongest transverse field areas exhibit a prominent alternating bright-dark spiral structure, which are different  from the usual appearance of LBs such as those shown by Wang et al. (2018). Notably, the observed spiral filaments are not parallel to the PIL as is usually seen in the regular LBs, and implied by Yang et al. (2017) for strong shear along PIL. In Figure 1(d),  we show a selected NIRIS Stokes U profile in the northern boxed region, where Hinode/SP level 2 data are saturated at 5000~G. Three Zeeman components are clearly seen. A direct measurement of the 1.9\AA~ Zeeman splitting ($\Delta\lambda=4.67\times10^{-13}g\lambda^2B$) yields a field strength B of 5570~G, about 1300~G above what Zirin \& Wang (1993) found for the strongest fields in flare productive $\delta$ sunspots.

Combining the very strong transverse field at the PIL and the spiral photospheric structure,  it is natural to ask:  are they signatures of a twisted magnetic flux rope right at the photospheric level?  Significant efforts will be devoted to further improve Stokes inversion of the Hinode/SP, SDO/HMI, and GST/NIRIS data. Okamoto \& Sakurai (2018) suggested the possibility that formation of a strong field may be due to the compression of fields at the two sides of a PIL.  Recent numerical simulations by Toriumi \& Takasao (2017) revealed that the strongly sheared PIL of $\delta$ sunspots is created by combined action of advection, stretching, and compression of magnetic fields.   These will motivate further analyses of the data and comparison with simulation to explain both the strong fields and the spiral LB structure of this AR. Furthermore,  these unusual behaviors may be related to its high productivity of major flares.

\begin{figure}
\epsscale{1.0}
\plotone{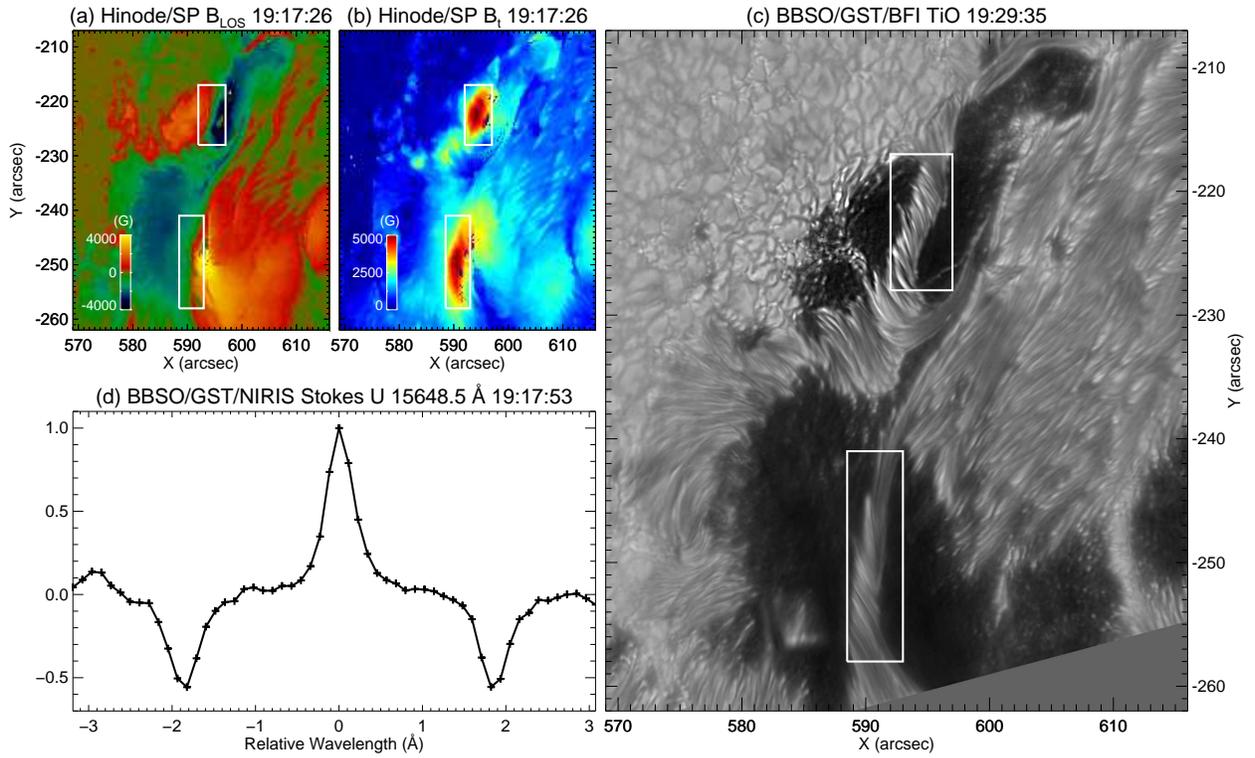}
\caption{Unusual structures of AR 12673. (a) and (b) Hinode/SP LOS and transverse magnetic field strength. Note that in many pixels near the PIL,  transverse fields are saturated at 5000~G. (c) BBSO/GST TiO image. The two white boxed in (a)--(c) mark the two strong transverse field areas at the PIL, where twisted photospheric LB structures of the $\delta$ configuration are present. (d) NIRIS Stokes U profile of a selected strong transverse field pixel at the PIL within the northern box. The direct measurement of Zeeman splitting yields a field strength of 5570~G. \label{f1}}
\end{figure}

\end{document}